\documentclass[twocolumn,showpacs,preprintnumbers,amssymb]{revtex4}
\usepackage{graphicx, epsfig, amssymb} %include figure files
\usepackage{amsmath, amsfonts}
\usepackage{bm}
\usepackage[usenames]{color}
\usepackage[
      colorlinks=true,
      linkcolor=blue,
      urlcolor=blue,
      filecolor=blue,
      citecolor=red,
      pdfstartview=FitV,
      pdftitle={},
      pdfauthor={},
      pdfsubject={},
      pdfkeywords={},
      pdfpagemode=None,
      bookmarksopen=true
]{hyperref}

\def\be{\begin{equation}}
\def\ee{\end{equation}}
\def\beq{\begin{eqnarray}}
\def\eeq{\end{eqnarray}}
\def\gsim{\mathrel{\rlap{\lower4pt\hbox{\hskip1pt$\sim$}}\raise1pt\hbox{$>$}}} 

\def\cM{{\cal M}}
\def\cJ{{\cal J}}
\def\cQ{{\cal Q}}

\begin{document}

\centerline{}
%\vskip 1cm
\title{Flowing along the edge: spinning up black holes in AdS spacetimes with test particles}

\author{
Jorge V. Rocha,$^{1}$
\footnote{Electronic address: jorge.v.rocha@tecnico.ulisboa.pt}
Raphael Santarelli,$^{2}$
\footnote{Electronic address: santarelli@ifsc.usp.br}
}
\affiliation{${^1}$ CENTRA,~Dept.~de F\'{\i}sica,~Instituto~Superior~T\'ecnico, Universidade de Lisboa, Av.~Rovisco Pais 1, 1049 Lisboa, Portugal}
\affiliation{${^2}$ Instituto de F\'isica de S\~ao Carlos, Universidade de S\~ao Paulo, Caixa Postal 369, CEP 13560-970, S\~ao Carlos, SP, Brazil}

\date{\today}

\begin{abstract}
We investigate the consequences of throwing point particles into odd-dimensional Myers-Perry black holes in asymptotically anti-de Sitter (AdS) backgrounds. We restrict our attention to the case in which the angular momenta of the background geometry are all equal. This process allows us to test the generalization of the weak cosmic censorship conjecture to asymptotically AdS spacetimes in higher dimensions. We find no evidence for overspinning in $D=5,7,9$ and $11$ dimensions. Instead, test particles carrying the maximum possible angular momentum that still fall into an extremal rotating black hole generate a flow along the curve of extremal solutions.
\end{abstract}


\maketitle

%\tableofcontents

%\clearpage

%%%%%%%%%%%%%%%%%%%%%%%%%%%%%%%%%%%%%%%%%%%
%%%%%%%%%%%%%%%%%%%%%%%%%%%%%%%%%%%%%%%%%%%
\section{Introduction}
\label{sec:Intro}

From the mathematical point of view, naked singula\-rities are just as valid as black holes (BHs), if regarded as solutions of General Relativity (GR).  Nevertheless, the appearance of a naked singularity as the endpoint of the evolution of regular and generic initial data would configure a `breakdown of predictability' within GR.  To avoid such a menace to the self-consistency of the theory, Penrose proposed back in 1969~\cite{Penrose:1969pc} what is now known as the (weak) cosmic censorship conjecture: any naked singularities formed in such a process always appear cloaked by an event horizon.

Despite many efforts over several decades, cosmic censorship remains unproven. This state of affairs naturally fostered attempts to disprove the conjecture by devising counter-examples. The first tests of the cosmic censorship were conducted by Wald~\cite{Wald:1974} and consisted of attempts to overspin/overcharge four-dimensional BHs with point particles.  The family of rotating and charged black holes --- known as Kerr-Newman solutions --- is described by mass $\cM$, angular momentum $\cJ$ and electric charge $\cQ$. The existence of an event horizon covering the central curvature singularity imposes a bound on these parameters,
\be
\cM \geq \sqrt{\cJ^2/\cM^2 + \cQ^2}\,.
\label{eq:bound}
\ee
This inequality is saturated for extremal BHs and if the bound is violated one obtains a naked singularity.
By considering an initially extremal Kerr-Newman black hole, Wald showed that point particles with enough angular momentum (charge) to overspin (overcharge) the BH are never captured, thus rendering such an attempt to disprove cosmic censorship unsuccessful.

Some years later Hubeny~\cite{Hubeny:1998ga} relaxed the extremality condition and demonstrated that an initially \textit{near-extremal} Reissner-Nordstr\"om BH could be overcharged by a test particle. In Refs.~\cite{Jacobson:2009kt, Saa:2011wq} the authors obtained similar results considering initially \textit{near-extremal} Kerr and Kerr-Newman BHs respectively. The validity of the cosmic censorship conjecture has also been investigated in higher dimensions~\cite{BouhmadiLopez:2010vc}, with a cosmological constant~\cite{Rocha:2011wp,Zhang:2013tba} or considering quantum tunneling effects~\cite{Matsas:2007bj,Hod:2008zza,Matsas:2009ww,Richartz:2011vf}. There are however indications that backreaction effects, which are typically neglected in studies reporting violation of cosmic censorship, restore the integrity of the BH horizon~\cite{Barausse:2011vx,Zimmerman:2012zu}.

In this work we perform numerical tests of the cosmic censorship --- in the spirit of Wald's classic gedanken experiment --- with rotating higher dimensional BHs with a negative cosmological constant, the so-called Myers-Perry-AdS spacetimes.
An extremal bound analogous to~\eqref{eq:bound} also exists for such solutions, but in this case it depends on the ratios between the several possible spin parameters.
We take advantage of the fact that, in odd dimensions ($D\geq5$), when all angular momenta are equal --- and we restrict to this case --- the solution is cohomogeneity-1~\cite{Kunduri:2006qa}. Thus the problem depends on a single (radial) coordinate, which amounts to a technical simplification.
We will present in detail the five-dimensional case but we also report our results for higher odd dimensions, specifically for $D=7,9,11$.

Besides being an interesting mathematical problem on its own right, the investigation of the cosmic censorship in higher dimensions is strongly motivated by recent numerical evidence in favor of its violation in five dimensions~\cite{Lehner:2010pn}.
The inclusion of a cosmological constant brings about added interest. From the viewpoint of the AdS/CFT correspondence, an over-extremal BH in AdS translates into a state on the boundary field theory rota\-ting faster than the speed of light~\cite{Hawking:1998kw}. Furthermore, although the ana\-lysis of a ring of test particles infalling into the three-dimensional rotating BH in AdS did not reveal any sign of cosmic censorship violation~\cite{Rocha:2011wp}, it was shown in Ref.~\cite{Mann:2008rx} --- by constructing exact thin-shell solutions --- that in three dimensions cosmic censorship can be violated if the matter on the shell has non-vanishing pressure.

The outline of the paper is the following. In Section~\ref{sec:ES-MP-AdS} we briefly present the black hole geometries we shall consider. Section~\ref{sec:conserv} discusses the geodesic motion of the test particles in these spacetimes and their conserved charges. Section~\ref{sec:spinup} includes the spin-up analysis of equal angular momenta MP-AdS geometries by test-particles. We conclude in Section~\ref{sec:conc} with some discussion and remarks.

%%%%%%%%%%%%%%%%%%%%%%%%%%%%%%%%%%%%%%%%%%%
%%%%%%%%%%%%%%%%%%%%%%%%%%%%%%%%%%%%%%%%%%%
\section{The background geometry
\label{sec:ES-MP-AdS}}

We will attempt to overspin higher dimensional rota\-ting black holes with a negative cosmological constant.
These solutions, which generalize the well known Myers-Perry family~\cite{Myers:1986un}, were first presented in~\cite{Hawking:1998kw} and have been extended to dimensions higher than five in~\cite{Gibbons:2004js}.
%In five spacetime dimensions one can pick two ortho\-gonal planes of rotation and in general the solution is parametrized by two independent angular momenta $a_1$ and $a_2$, in addition to a mass parameter $M$.
In odd $D$ dimensions, with $D \equiv 2N+3 \geq 5$, there are $N+1$ independent rotation planes and, as shown in~\cite{Kunduri:2006qa}, when the rotation parameters are set all equal, $a_i=a$, the isometry group of these BH spacetimes is enhanced and a system of coordinates can be found such that the metric only depends on a single radial  coordinate,
\beq
 ds^2  &=& g_{\mu\nu}dx^\mu dx^\nu = - f(r)^2 dt^2  + g(r)^2 dr^2 \nonumber \\
  && \!\!\!\!\!\!\!\!\!\!\!\!\!\!\!\! + r^2 \widehat{g}_{ab} dx^a dx^b +\, h(r)^2 \left[ d\psi + A_a dx^a - \Omega(r) dt \right]^2,
\label{eq:metric}
\eeq
where
\beq
g(r)^2  &=&  \left( 1 + \frac{r^2}{\ell^2} - \frac{2M\Xi}{r^{2N}} + \frac{2Ma^2}{r^{2N+2}} \right)^{-1}\,, \\
h(r)^2  &=&  r^2 \left( 1 + \frac{2Ma^2}{r^{2N+2}} \right)\,, \qquad \Omega(r) =  \frac{2Ma}{r^{2N} h(r)^2}\,, \\
f(r)  &=&  \frac{r}{g(r) h(r)}\,, \qquad \Xi = 1 - \frac{a^2}{\ell^2}\,.
\eeq
In writing these expressions we are using coordinates $x^\mu=\{t,r,\psi,x^a\}$, $a=1,\dots,2N$. The $x^a$ are (real) coordinates parametrizing the base space, which turns out to be the complex projective space $CP^N$. The appearance of this particular base space is a consequence of the fact that a sphere $S^{2N+1}$ can be written as an $S^1$ bundle over $CP^N$. The coordinate $\psi$ parametrizes the $S^1$ fiber and has period $2\pi$. In Eq.~\eqref{eq:metric}, $\widehat{g}_{ab}$ denotes the Fubini-Study metric on $CP^N$ and $A=A_a dx^a$ is its K\"ahler potential. Explicit expressions for both $\widehat{g}_{ab}$ and $A$ can be constructed iteratively (see for example~\cite{Dias:2010eu}), though we shall not require them for our study.

In the simplest case, $D=5$, the base space is $CP^1$, which is isomorphic to the sphere $S^2$, and
\be
\widehat{g}_{ab} dx^a dx^b  =  \frac{1}{4} \left( d\theta^2 + \sin^2\theta \, d\phi^2  \right)\,, \;
A = \frac{1}{2} \cos\theta \, d\phi\,.
\ee
Defining $\theta\equiv x^1$ and $\phi\equiv x^2$, the two orthogonal rotation planes correspond to $\theta=0$ and $\theta=\pi$ in these coordinates, i.e., the rotation planes are mapped to the poles of the $S^2$. For generic $N\geq1$ we shall group the base space coordinates as $x^a = (\theta^i,\phi^i)$, with $i=1,\dots,N$.

The metric~\eqref{eq:metric} is a solution of the Einstein equations with a negative cosmological constant,
\be
R_{\mu\nu} = - (D-1)\ell^{-2} g_{\mu\nu}\,.
\ee
The largest real root, $r_+$, of $g^{-2}$ marks an event horizon which possesses the geometry of a homogeneously squashed $S^{2N+1}$.
The mass ${\cal M}$ and angular momentum ${\cal J}$ of the spacetime are given by~\cite{Kunduri:2006qa}
\beq
{\cal M} &=& \frac{\Omega_{2N+1}}{4\pi G} M \left( N + \frac{1}{2} + \frac{a^2}{2\ell^2} \right)\,, \label{eq:mass}\\
{\cal J} &=& \frac{\Omega_{2N+1}}{4\pi G} (N+1) M a \,, \label{eq:angmom}
\eeq
where $\Omega_{2N+1}$ is the area of the unit $(2N+1)-$sphere.
There exists a regular extremal solution for which the horizon becomes degenerate, corresponding to a maximal value of $a$ for given $M$ and $\ell$.
However, for non-vanishing cosmological constant this bound cannot be expressed in closed form.
%this bound is more easily expressed in terms of the angular {\em velocity} of the horizon, $\Omega_H$, and the location of the event horizon, $r_+$~\cite{Kunduri:2006qa}:
%
%\be
%\Omega_H = \frac{2Ma}{r_+^{2N+2} + 2Ma^2} \leq \frac{1}{\ell} \sqrt{1+\frac{N\ell^2}{(N+1)r_+^2}}\,.
%\ee
%

It will be convenient to employ a dimensionless radial coordinate $\rho \equiv r^2/\ell^2$ as well as dimensionless combinations for mass and spin, namely
\be
m \equiv \frac{M}{\ell^{2N}} \,, \qquad j \equiv \frac{a}{\sqrt{M}}\ell^{N-1}\,,
\label{eq:dimlessparams}
\ee
in terms of which the metric function $g(r)$ becomes
\be
g(\rho)^{-2} = 1+\rho +2 m \rho ^{-(N+1)} \left[ j^2 m +\rho (j^2 m - 1) \right]\,.
\label{eq:dimlessg}
\ee

In the rest of the manuscript we will use geometrical units with $G=c=1$.

%%%%%%%%%%%%%%%%%%%%%%%%%%%%%%%%%%%%%%%%%%%
\section{Conserved quantities and geodesics
\label{sec:conserv}}

The stationary spacetimes under consideration possess both timelike and rotational Killing vectors. The me\-tric tensor $g_{\mu\nu}$ is explicitly independent of coordinates $\{t,\psi,\phi^i\}$. This immediately gives $N+2$ conserved quantities:
\begin{flalign}
E &\equiv -g_{\mu\nu} (\partial_t)^\mu \dot z^\nu = - \left[  g_{tt}\dot{T} +g_{t\psi}\dot{\Psi}+g_{t \phi^j}{\dot{\Phi}^j} \right], \label{eq:EandL1}\\
L_\psi &\equiv g_{\mu\nu} (\partial_\psi)^\mu \dot z^\nu = g_{t\psi}\dot{T}+g_{\psi\psi}\dot{\Psi}+g_{\psi \phi^j}{\dot{\Phi}^j}, \label{eq:EandL2}\\
L_{\phi^i} &\equiv g_{\mu\nu} (\partial_{\phi^i})^\mu \dot z^\nu = g_{t \phi^i}\dot{T}+g_{\psi \phi^i}\dot{\Psi}+g_{\phi^i \phi^j}{\dot{\Phi}^j}, \label{eq:EandL3}
\end{flalign}
where $z^\mu(\tau)=(T(\tau),R(\tau),\Psi(\tau),X^a(\tau))$ stands for the particle coordinates and the dot indicates differentiation with respect to the affine parameter $\tau$.

Let us first particularise to $N=1$ for concreteness. Note that
\beq
g_{t\phi} &=& \frac{\cos\theta}{2} g_{t\psi}\,, \qquad  g_{\psi\phi} = \frac{\cos\theta}{2} g_{\psi\psi}\,, \nonumber\\
g_{\phi\phi} &=& \frac{\cos\theta}{2} g_{\psi\phi} + \frac{r^2 \sin^2\theta}{4}\,,
\eeq
so a test particle whose motion lies entirely in the rotation plane $\theta=0$ has $L_\phi = \frac{1}{2} L_\psi$.
Similarly, a test particle whose motion lies entirely in the rotation plane $\theta=\pi$ has $L_\phi = -\frac{1}{2} L_\psi$.
These geodesics have
\be
\dot{\Theta}(\tau) = \dot{\Phi}(\tau) = 0\,.
\label{eq:geodesics}
\ee
In fact, there exist geodesics obeying~\eqref{eq:geodesics} {\em for any value of $\theta$ and $\phi$}.
Such geodesics simply correspond to static trajectories on the $S^2$ and they satisfy
\be
L_\phi= \frac{\cos\theta}{2}L_\psi \,,
\ee

Now consider the general case $N\geq1$. It is easy to see from~\eqref{eq:metric} that
\be
g_{t \phi^i} = A_{\phi^i} g_{t\psi}\,, \qquad  g_{\psi \phi^i} = A_{\phi^i} g_{\psi\psi}\,.
\label{eq:relategmunu}
\ee
The geodesics we shall consider for the test particles are the generalization of Eq.~\eqref{eq:geodesics},
\be
\dot{X^a}(\tau) = 0\,,
\label{eq:geodesicsN}
\ee
which implies in particular $\dot{\Phi}^j(\tau)=0$, and therefore sa\-tisfy 
\be
L_{\phi^i}= A_{\phi^i} L_\psi \,,
\ee
which follows from Eqs.~\eqref{eq:EandL2}, \eqref{eq:EandL3} and~\eqref{eq:relategmunu}.

The remaining equations governing the test particle trajectories are easily obtained by inverting~\eqref{eq:EandL1} and~\eqref{eq:EandL2} and using $g_{\mu\nu}\dot z^\mu \dot z^\nu=-\epsilon$ for the radial equation:
\beq
\dot{T}     &=& \frac{E- \Omega L_\psi}{f^2}\,,\\
\dot{\Psi}  &=& \frac{h^2\Omega E + \left( f^2-h^2\Omega^2 \right)L_\psi}{h^2f^2}\,,\\
\dot{R}^2   &=& - \frac{ \epsilon - f^{-2}\left[ E-\Omega L_\psi \right]^2 +h^{-2}L_\psi^2}{g^2}\,,\label{eq:radialequation} 
\eeq
where $\epsilon=1,0$ for timelike or null geodesics, respectively.
In these equations the metric functions $f,g,h,\Omega$ should all be understood as functions of $R(\tau)$ instead of $r$.

One final comment is in order before conducting the test of the cosmic censorship in the next section.
It is not obvious how to define separately the `black hole' and the `test particle' when the latter cannot be moved all the way out to timelike infinity. Therefore, for the spin-up analysis we will restrict to the case $\epsilon=0$, since only null geodesics can extend all the way out to the boundary of AdS.

%%%%%%%%%%%%%%%%%%%%%%%%%%%%%%%%%%%%%%%%%%%
%%%%%%%%%%%%%%%%%%%%%%%%%%%%%%%%%%%%%%%%%%%
\section{Spinning-up AdS black holes with test particles
\label{sec:spinup}}

In order to test the cosmic censorship conjecture in a higher dimensional and asymptotically AdS setting we will now attempt to spin up {\em extremal} Myers-Perry-AdS black holes with all angular momenta equal.

We thus take, as our background, solutions of the form~\eqref{eq:metric} with initial mass ${\cal M}_0$ and angular momentum ${\cal J}_0$ corresponding to an extremal configuration with $j=j_{ext}(m)$. Note that the extremal value of the dimensionless spin parameter $j$ depends on the dimensionless mass parameter $m$. This is in contrast with the case of asymptotically flat black holes~\cite{BouhmadiLopez:2010vc}, for which the extremal bound --- with a different definition of the parameter $j$ --- turned out to be a (dimension-dependent) constant. This difference is intimately related with the fact that for a non-vanishing cosmological constant the physical mass of the spacetime depends explicitly on the spin parameter $a$ [see eq.~\eqref{eq:mass}].

Next, we try to overspin these black holes by throwing in $(D-1)/2= N+1$ identical particles, one along each orthogonal rotation plane, in order to preserve the equal angular momenta property of the background.
All test particles have the same mass $\mu/(N+1)$ and the same conserved gravitational charges. If, as an outcome of the gedanken experiment, these particles fall past the event horizon, they will induce a change in mass and angular momentum of the black hole given respectively by
\be
\delta\,{\cal M} = \mu E\,, \qquad
\delta\,{\cal J} = \mu L_\psi\,.
\label{eq:variations1}
\ee
These must satisfy $\delta{\cal M}\ll {\cal M}_0$ and $\delta{\cal J}\ll {\cal J}_0$ for the test particle limit to be valid. We thus regard $\mu$ as an infinitesimal quantity and present our final expressions as series expansions around $\mu=0$ truncated at first order.

The right hand side of Eqs.~\eqref{eq:variations1} conforms with the energy and angular momentum carried by point particles in asymptotically flat spacetimes, for which asymptotically there is no gravitational potential. It is not immediate that the same expressions apply in the asymptotically AdS case. Nevertheless, one can compute the effect the introduction of a test particle in the background geometry has on the conserved charges of the spacetime. This was done in~\cite{Delsate:2014} and the results are consistent with the identification of the parameter $E$ with `energy density' and the parameter $L_\psi$ with `angular momentum density'.

%%%%%%%%%%%%%%%%%%%%%%%%%%%%%%%%%%%%%%%%%%%
\subsection{Extremal equal angular momenta black holes
\label{sec:extremalBHs}}

As mentioned above, we must first identify the class of extremal solutions among the family of equal angular momenta Myers-Perry-AdS black holes in odd spacetime dimensions. This can be done analytically for asymptotically flat BHs but in the presence of a non-vanishing cosmological constant even this first step must be performed numerically. In any case, it amounts to determining the value of the dimensionless spin parameter $j$ (as a function of $m$) for which the event horizon is degenerate with the inner Cauchy horizon. In other words, the extremal limit occurs when the metric function $g(\rho)^{-2}$ in Eq.~\eqref{eq:dimlessg} features a double root for a positive value of $\rho$, which corresponds to the 
horizon of the extremal black hole, $\rho_{hor}$.

\begin{figure}%[htb!]
\centering
\includegraphics[width=8.6cm]{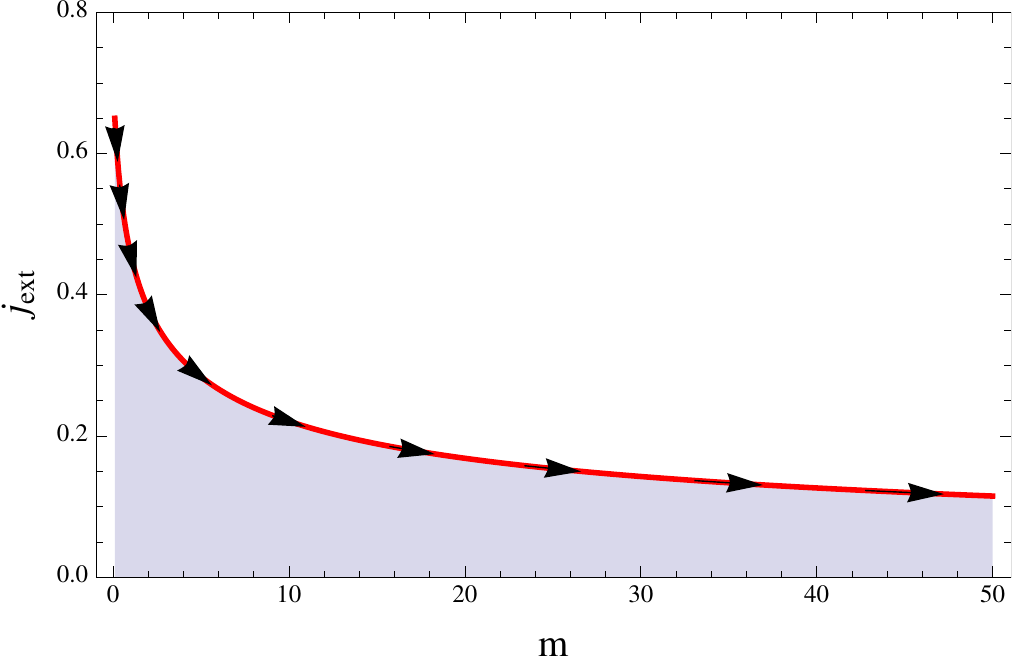}
\caption{Curve of extremal, equally rotating Myers-Perry-AdS black holes in parameter space $(m,j)$, for $D=5$. Black holes with regular event horizons exist only in the shaded region below the extremal curve (solid red). The arrows indicate the direction (conveniently normalised) of the bidimensional vector $(\delta m, \delta j)$, as obtained in Sec.~\ref{sec:CCCtest}. The spin-up process conducted with extremal black holes and the potentially most dangerous geodesics yields a flow {\em along} the family of extremal solutions.}
\label{fig:j_ext}
\end{figure}

In Figure~\ref{fig:j_ext} we present our results for the extremal curve in the case $D=5$, obtained by imposing the conditions described above for values of $m$ covering the interval $[0.1, 50]$ in steps of 0.1.
Figure~\ref{fig:rho} shows the radial location of the horizon for such 5D extremal black holes.

The infinite $D$ limit of the extremal curve can nevertheless be obtained analytically. It follows straightforwardly from imposing the vanishing of Eq.~\eqref{eq:dimlessg} and of its derivative at $\rho=\rho_{hor}$ and then taking $N\to\infty$. The result for the extremal curve and its derivative is
\be
j_{ext}(m) \stackrel{N\to\infty}{\longrightarrow} \frac{1}{\sqrt{2m}}\,, \qquad
j_{ext}'(m) \stackrel{N\to\infty}{\longrightarrow} -\frac{m^{-3/2}}{2\sqrt{2}}\,.
\label{eq:infiniteN}
\ee
%

%%%%%%%%%%%%%%%%%%%%%%%%%%%%%%%%%%%%%%%%%%%
\subsection{Critical geodesics
\label{sec:xcrit}}

The next step is to determine the critical value of the test particles angular momentum parameter $L_\psi$ (for given energy perameter $E$) above (below) which the geodesics correspond to a bounce (plunge) in the extremal black hole background.

Equation~\eqref{eq:radialequation} specifies the radial motion of the test-particles and we are interested in determining for which values of the pair $(E,L_\psi)$ the particle is absorbed by the black hole. Considering only null geodesics, $\epsilon=0$, we have
\be
\dot R^2 + V(R) = 0\,,
\ee
with the effective potential for radial motion given by
\beq
V(R) &\equiv& - \left(1 + \frac{2 a^2 M}{R^{2(N+1)}}\right)E^2 + \frac{4 a M \ell}{R^{2(N+1)}} E \frac{L_{\psi}}{\ell}  \nonumber\\ 
&& + \frac{R^{2(N+1)}+\ell^2 R^{2N} -2 M \ell^2}{R^{2(N+1)}} \frac{L_{\psi}^2}{\ell^2}\,.
\eeq

It is clearly convenient to introduce a parameter $x$ defined as
\be
L_\psi \equiv x E \ell\,.
\ee
Then, the effective radial potential in terms of the new radial coordinate $\rho=r^2/\ell^2$, the dimensionless para\-meters defined in Eq.~\eqref{eq:dimlessparams} and the parameter $x$ defined above, simplifies:
\be
V(\rho) = - \frac{E^2}{\rho^{N+1}} \left[ 2m \left(j \sqrt{m} - x\right)^2 + \rho^N \left(\rho(1 -x^2) - x^2 \right) \right]\,.
\ee
Turning points in this geometry thus correspond to real roots of the effective potential $V$, when located outside the event horizon.

\begin{figure}%[htb!]
\centering
\includegraphics[width=8.6cm]{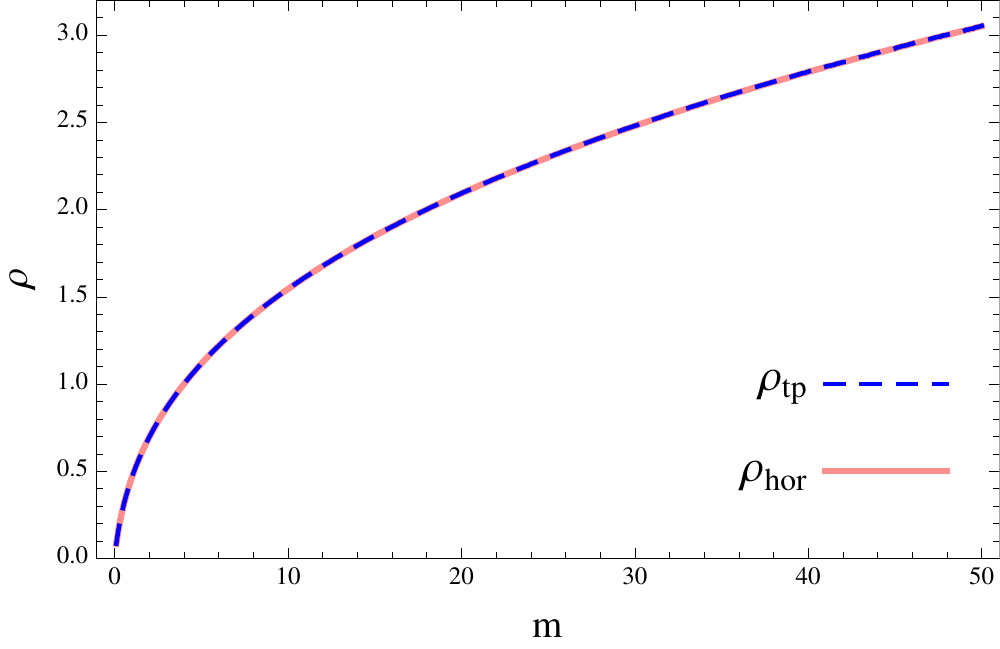}
\caption{Radial location, in dimensionless coordinate $\rho$, of the event horizon (solid pink line) for the class of extremal black holes in $D=5$ spacetime dimensions, as a function of the dimensionless mass parameter $m$. Superimposed is shown (blue dashed line) the location of the turning point for the critical geodesics. The two curves match, up to numerical precision.}
\label{fig:rho}
\end{figure}

\begin{figure}%[htb!]
\centering
\includegraphics[width=8.6cm]{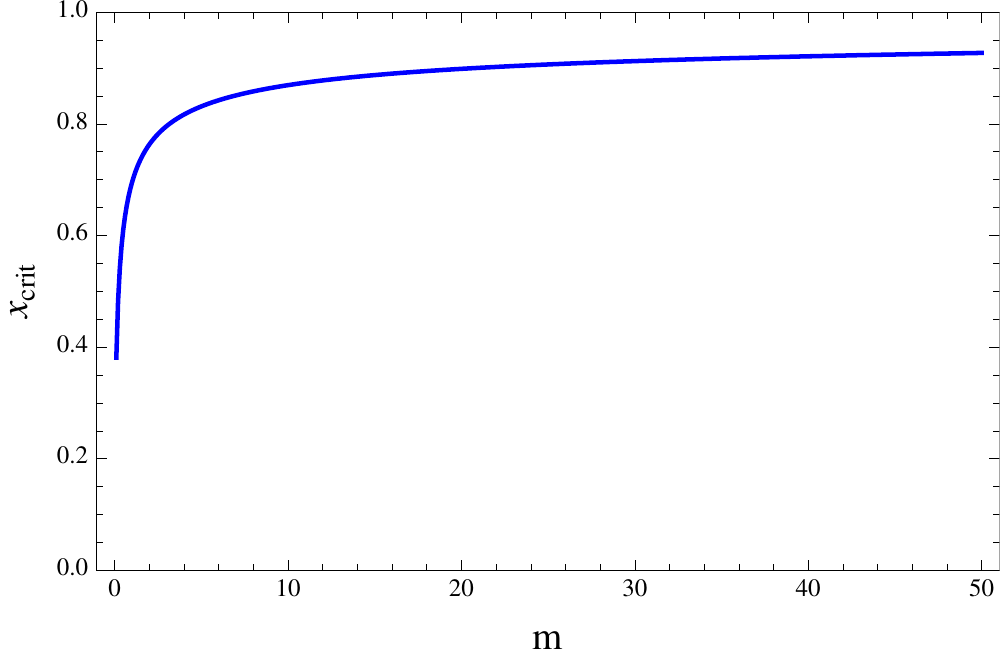}
\caption{Critical value of the dimensionless parameter $x$ as a function of the dimensionless mass parameter $m$, for the class of extremal, equal angular momentum black holes in $D=5$ spacetime dimensions. Geodesics with $x<x_{crit}$ are absorbed while geodesics with $x>x_{crit}$ correspond to bounces.}
\label{fig:xcrit}
\end{figure}

Clearly, the situation offering the best chances of overspinning the black hole is when the initial geometry is extremal and the particles' angular momentum attains the maximum value for which absorption still occurs. We denote the $x$ parameter in this situation by $x_{crit}$ and this quantity as a function of $m$ (for $j=j_{ext}(m)$) is shown in Fig.~\ref{fig:xcrit}.
Furthermore, our numerical results also indicate that such critical geodesics have a turning point that precisely coincides with the event horizon (see Fig.~\ref{fig:rho}).

%%%%%%%%%%%%%%%%%%%%%%%%%%%%%%%%%%%%%%%%%%%
\subsection{Testing cosmic censorship
\label{sec:CCCtest}}

Finally, we need to interpret our results to infer whether or not cosmic censorship conjecture is violated by Wald's thought experiment in this higher-dimensional, asymptotically AdS context. This is not as immediate as in Ref.~\cite{BouhmadiLopez:2010vc} because the curve of extremal solutions $j_{ext}(m)$ depends on $m$ and this dependence is only known nume\-rically. In general, the absorption of test-particles by the black hole will produce a change in the dimensionless spin parameter $\delta j$, as well as on the dimensionless mass parameter, $\delta m$. 

Upon absorption of the test particles, the dimensionless mass and spin of the BH, defined in Eq.~\eqref{eq:dimlessparams}, change according to
\be
m_0 \to m_0+\delta m\,, \qquad
j_0 \to j_0+\delta j\,,
\ee
where the subscript stands for initial parameters of the BH, which we take to be extremal, i.e., $j_0=j_{ext}(m_0)$.
The expressions for $\delta m$ and $\delta j$ are of course obtained by varying Eqs.~\eqref{eq:dimlessparams} and using~\eqref{eq:variations1}:
\begin{widetext}
\beq
\delta m &=& \frac{\partial m}{\partial M} \delta M + \frac{\partial m}{\partial a} \delta a
= \left( \frac{\partial m}{\partial M} \frac{\partial M}{\partial {\cal M}} + \frac{\partial m}{\partial a} \frac{\partial a}{\partial {\cal M}} \right) \delta {\cal M}
+ \left( \frac{\partial m}{\partial M} \frac{\partial M}{\partial {\cal J}} + \frac{\partial m}{\partial a} \frac{\partial a}{\partial {\cal J}} \right) \delta {\cal J} \nonumber\\
&=& \frac{2\mu E }{\omega \ell^{2N} \left(2N+1-j^2 m\right)} \left[1-\frac{j \sqrt{m}}{N+1}x\right]\,, \\
\delta j &=& \frac{\partial j}{\partial M} \delta M + \frac{\partial j}{\partial a} \delta a
= \left( \frac{\partial j}{\partial M} \frac{\partial M}{\partial {\cal M}} + \frac{\partial j}{\partial a} \frac{\partial a}{\partial {\cal M}} \right) \delta {\cal M}
+ \left( \frac{\partial j}{\partial M} \frac{\partial M}{\partial {\cal J}} + \frac{\partial j}{\partial a} \frac{\partial a}{\partial {\cal J}} \right) \delta {\cal J} \nonumber\\
&=& \frac{j\mu E}{m\omega \ell^{2N}\left(2N+1-j^2 m\right)} \left[\frac{2N+1+2 j^2 m}{(N+1)j\sqrt{m}} x-3\right]\,.
\label{eq:variations2}
\eeq
\end{widetext}

As can be seen in Fig.~\ref{fig:j_ext}, even if the variation $\delta j$ turns out to be ne\-gative, a positive $\delta m$ might lead to a situation in which 
\be
j_{ext}(m_0) + \delta j > j_{ext}(m_0 + \delta m)\,,
\ee
and such an occurrence would signal the overspinning of an extremal black hole.
Therefore, the accurate condition for violation of cosmic censorship is not that $\delta j>0$, but instead
\be
\frac{\delta j}{\delta m} > j_{ext}'(m)\,,
\ee
i.e., it is the {\em direction} of the bidimensional vector $(\delta m, \delta j)$ that must be compared with the {\em slope} of the extremal curve $j_{ext}(m)$.

\begin{figure}[t!]%[htb!]
\centering
\includegraphics[width=8.6cm]{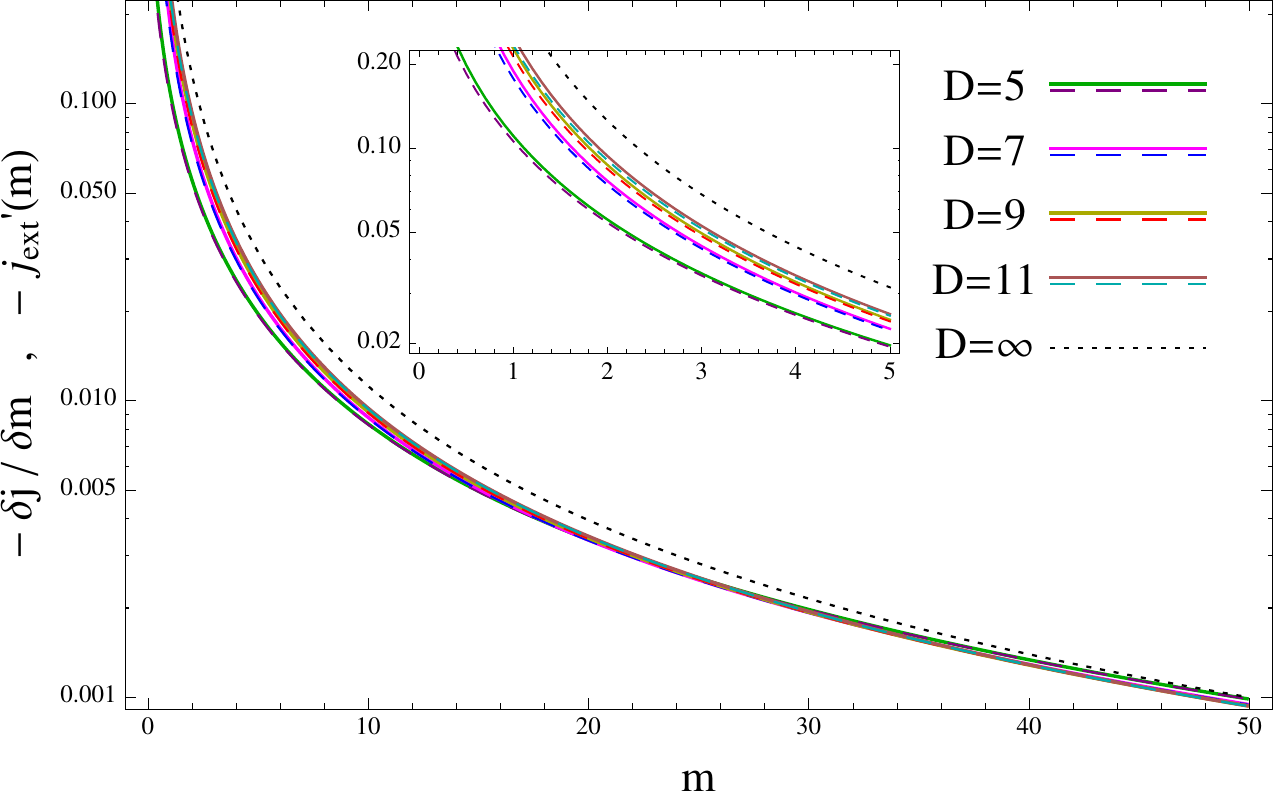}
\caption{Comparison between the slope of the bidimensional vector $(\delta m, \delta j)$ (solid lines) and the derivative of the extremal bound, $j_{ext}'(m)$ (dashed lines), for $D=5,7,9,11$. Both quantities are negative and our numerical computations reveal $\delta j/\delta m$ is everywhere (slightly) more negative than $j_{ext}'(m)$, as is necessary to satisfy cosmic censorship. Nevertheless, the two curves match within numerical precision. The dotted line represents the analytic result for the curve $j_{ext}'(m)$ in the limit $D\to\infty$, according to Eq.~\eqref{eq:infiniteN}.
%Note that a logarithmic scale is used for the ordinate axis.
}
\label{fig:djdm}
\end{figure}

These two quantities are plotted in Fig.~\ref{fig:djdm} and the results show no indication of cosmic censorship violation in $D=5,7,9$ and $11$ dimensions. In fact, in the region of parameter space where our computations attain better resolution ($m\gsim 1$) our results are fully consistent with $\delta j/\delta m = j_{ext}'(m)$, yielding a relative difference of the order of $1\%$ or less. For smaller values of the dimensionless mass parameter $m$ the relative difference is somewhat larger but this is due to lack of resolution in the $m$ parameter.

The attempt to spin-up extremal backgrounds with the potentially most dangerous geodesics --- those with $x=x_{crit}$ --- is on the verge of overspinning the black holes. Instead, the process generates a flow {\em along} the family of extremal solutions, as shown in Fig.~\ref{fig:j_ext} for $D=5$. Similar results were obtained for the other odd dimensions analysed.

%%%%%%%%%%%%%%%%%%%%%%%%%%%%%%%%%%%%%%%%%%%
%%%%%%%%%%%%%%%%%%%%%%%%%%%%%%%%%%%%%%%%%%%
\section{Discussion
\label{sec:conc}}

In summary, we have performed tests of the weak cosmic censorship --- in the spirit of Wald's classic {\it gedanken} experiment --- on asymptotically AdS rotating black holes possessing equal angular momenta, for $D=5,7,9$ and $11$ spacetime dimensions. We observed no violation of the conjecture.
More interestingly, our results indicate that starting with an initially extremal BH and throwing in test particles that (i) preserve the equal angular momentum property of the background and (ii) carry the maximum allowed angular momentum for which absorption still occurs, one can indeed spin up the black hole but the outcome is a more massive BH which is also extremal.

In effect, the spin up process investigated originates a flow along the curve of extremal black holes. This is in line with the findings of Ref.~\cite{Wald:1974} for Kerr-Newman black holes and of Ref.~\cite{BouhmadiLopez:2010vc} for the $D=5$ asymptotically flat Myers-Perry solution, where a similar phenomenon was observed. It is therefore tempting to speculate about the generalization of this result to other, possibly all, black hole solutions, irrespective of their asymptotics. Based on the similarity of the results for the different dimensions analysed and the structure of the governing equations being the same, it is very likely that our conclusions apply to all odd higher dimensions as well. This is also supported by the relatively small difference between the extremal curve $j_{ext}'(m)$ for $D=11$ and for $D\to\infty$ [see Fig.~\ref{fig:djdm}].

Let us briefly comment on an apparent disagreement between our results and those of Ref.~\cite{Zhang:2013tba}, which also considered overspinning extremal Kerr-AdS black holes, though restricting to $D=4$ dimensions. In~\cite{Zhang:2013tba}, violation of cosmic censorship was reported to occur for certain fine-tunned values of the parameters, still within the test particle approximation. However, timelike geodesics in asymptotically AdS spacetimes always possess turning points outside the horizon. The physical meaning is clear: massive particles cannot be sent into a black hole in AdS from arbitrarily far away. Instead, we have considered only massless particles that can travel in from spatial infinity, and our previous conjecture can only apply in this situation.

Finally, recall that the background metrics we attempted to spin up were strictly extremal, as opposed to near-extremal, and the effect of the absorption of point particles was only considered at first order. This is in keeping with the test particle approximation employed, but given the delicate outcome of our results it is conceivable that higher order effects might alter the picture. As usual backreaction effects, such as gravitational radiation emission and self-force that have been neglected, are expected to weaken the possibility of overspinning the black holes.

%%%%%%%%%%%%%%%%%%%%%%%%%%%%%%%%%%%%%%%%%%%
%%%%%%%%%%%%%%%%%%%%%%%%%%%%%%%%%%%%%%%%%%%
\section*{Acknowledgements}

It is a pleasure to thank Antonino Flachi and Vincenzo Vitagliano for comments and a careful reading of a draft.
JVR is supported by {\it Funda\c{c}\~ao para a Ci\^encia e Tecnologia} (FCT)-Portugal through contract No.~SFRH/BPD/47332/2008. RS is supported by grant \#2012/20039-6, S\~ao Paulo Research Foundation (FAPESP). RS also thanks \textit{Centro Multidisciplinar de Astrof\'isica} and \textit{Instituto Superior T\'ecnico} (IST) for hospitality during the initial phase of this work.

%%%%%%%%%%%%%%%%%%%%%%%%%%%%%%%%%%%%%%%%%%%
%%%%%%%%%%%%%%%%%%%%%%%%%%%%%%%%%%%%%%%%%%%

\end{document}